\documentclass[aps,reprint,superscriptaddress,nofootinbib]{revtex4-2}

\usepackage{amsmath}
\usepackage{graphicx}
\usepackage{mathtools}
\usepackage{hyperref}
\hypersetup{colorlinks=true,linkcolor=black,citecolor=blue,urlcolor=blue}
\usepackage{enumitem} 

\begin{document}

\title{Quantum-statistical transport phenomena in memristive computing architectures}
%\title{Dynamic disorder: A unified quantum-statistical framework for memristive transport}
 
\author{Christopher N. Singh}
%\email[email: ]{csingh5@bighamton.edu}
%\thanks{Corresponding author}
%\homepage[visit webpage at: ]{http://www.linkedin.com/in/csingh5binghamton/}
\affiliation{Department of Physics, Applied Physics, and Astronomy, 
Binghamton University, Binghamton, New York 13902, USA}
\affiliation{Materials Science and Technology Division, Los Alamos National Laboratory, Los Alamos, NM 87545, USA}

\author{Brian A. Crafton}
\affiliation{Department of Electrical and Computer Engineering, Georgia
Institute of Technology, Atlanta, GA, 30332, USA}

\author{Mathew P. West}
\affiliation{Department of Electrical and Computer Engineering, Georgia
Institute of Technology, Atlanta, GA, 30332, USA}

\author{Alex S. Weidenbach}
\affiliation{Department of Electrical and Computer Engineering, Georgia
Institute of Technology, Atlanta, GA, 30332, USA}

\author{Keith T. Butler}
\affiliation{SciML, Scientific Computing Department, Rutherford Appleton Laboratory, Didcot, OX110QX, UK}

\author{Allan H. MacDonald}
\affiliation{Department of Physics, University of Texas at Austin, Austin Texas 78712-1081, USA}

\author{Arjit Raychowdury}
\affiliation{Department of Electrical and Computer Engineering, Georgia
Institute of Technology, Atlanta, GA, 30332, USA}

\author{Eric M. Vogel}
\affiliation{Department of Electrical and Computer Engineering, Georgia
Institute of Technology, Atlanta, GA, 30332, USA}

\author{W. Alan Doolittle}
\affiliation{Department of Electrical and Computer Engineering, Georgia
Institute of Technology, Atlanta, GA, 30332, USA}

\author{L. F. J. Piper}
%\email[email: ]{lpiper@binghamton.edu}
%\thanks{Corresponding author}
\affiliation{Department of Physics, Applied Physics, and Astronomy, 
Binghamton University, Binghamton, New York 13902, USA}
\affiliation{WMG, University of Warwick, Coventry CV4 7AL, UK}

\author{Wei-Cheng Lee}
%\email[email: ]{wlee@binghamton.edu}
%\thanks{Corresponding author}
\affiliation{Department of Physics, Applied Physics, and Astronomy, 
Binghamton University, Binghamton, New York 13902, USA}
%\homepage[visit webpage at: ]{http://bingweb.binghamton.edu/~wlee/} 

\date{\today}

\begin{abstract}
The advent of reliable, nanoscale memristive components is promising for next
generation compute-in-memory paradigms, however, the intrinsic variability in
these devices has prevented widespread adoption. Here we show coherent electron
wave functions play a pivotal role in the nanoscale transport properties of
these emerging, non-volatile memories. By characterizing both filamentary and
non-filamentary memristive devices as disordered Anderson systems, the
switching characteristics and intrinsic variability arise directly from the
universality of electron transport in disordered media. Our framework suggests
localization phenomena in nanoscale, solid-state memristive systems are
directly linked to circuit level performance. We discuss how quantum
conductance fluctuations in the active layer set a lower bound on device
variability. This finding implies there is a fundamental quantum limit on the
reliability of memristive devices, and electron coherence will play a decisive
role in surpassing or maintaining Moore's Law with these systems. 
\end{abstract}

\maketitle

\section{Introduction}
The von Neumann model of computing is the foundation of modern digital
technologies, but this paradigm harbors an intrinsic bottleneck. This
bottleneck arises because data cannot be operated on in the same place that it
is stored, and therefore, computation speed is limited by the rate data can be
transferred between memory and compute locations~\cite{williams1948electronic}.
Because of this, solid-state, bio-mimetic computing that avoids this inefficient
data transfer  and enables a `compute-in-memory' paradigm is highly
desired~\cite{sangwan2020neuromorphic}. However, mimicking biological systems that
typically have billions of neurons~\cite{indiveri2013integration} suggests a
solid-state analog would require a similar number of logical units. With
transistors already hitting the quantum limits of
performance~\cite{powell2008}, a different class of nanoscale, solid-state
components is needed, and adaptive-oxide memristors are considered among the most
promising candidates~\cite{waser2004nanoelectronics,yang2008}. Although viable
solid-state memristors have already been reported~\cite{zhu2020comprehensive}, the
variability inherent in their operation has been severely detrimental to
circuit-level performance~\cite{yang2009}.  What's more, their underlying
switching mechanisms have been hotly debated~\cite{sun2019understanding,
pickett2013, delvalle2017, waser2009}.  

In solid-state devices, the migration of atomic defects is generally considered
sufficient for memristive switching~\cite{strukov2008}, but whether it is
necessary--or even optimal--is unclear~\cite{wang2019overview}, as
purely-electronic switching mechanisms have also been
proposed~\cite{pickett2013,del2020caloritronics}. In either case, quantum
transport effects such as strong electronic correlations, interface scattering,
tunneling, and interference may all additionally contribute to the total
transport characteristics. Thus, to improve the performance of nanoscale,
compute-in-memory devices, a quantum-theoretical framework is necessary. This
is especially pressing given that predicting device properties post fabrication
remains the single greatest challenge to widespread
adoption~\cite{molas2018resistive,lian2019resistance}. Treating all the
relevant variables in a non-equilibrium, quantum framework will therefore
accelerate the development pipeline of memristive materials~\cite{panda2018}. 

This work fills that gap in the microscopic description of memristive transport
by developing a computationally-tractable, first-principles approach to predict
transport properties of electrons in environments with stochastic disorder
potentials.  We apply it to filamentary (a-HfO$_x$) and non-filamentary
(a-Nb$_2$O$_{5-x}$) systems. In filamentary devices, we find that
electron-phase effects are a significant contribution of variability.  In
non-filamentary systems, we find that dynamic disorder potentials can give rise
to hysteretic conductance curves.  Together, these results indicate that
electron-phase effects can significantly influence the performance of
compute-in-memory devices. The stochastic nature of the conductance in
filamentary systems~\cite{yi2016quantized} is shown to be analogous to that of
quantum wires near an Anderson localization transition~\cite{muttalib1999}.
One of the most astounding properties of quantum disordered systems is that the
logarithm of the conductance, not the resistance, stabilizes in the
thermodynamic limit~\cite{abrikosov1981paradox}.  In this way, the conductance
becomes a key circuit-level design element because near an Anderson transition,
the resistance is not self-averaging.  This immediately suggests interference
phenomena could play a central role in the rational design of bio-mimetic
hardware. We anticipate an understanding of the origin of variability will
enable circuit engineers to design more robust compute-in-memory architectures.

\begin{figure*}[ht!]
	\centering
	\includegraphics[width=\textwidth]{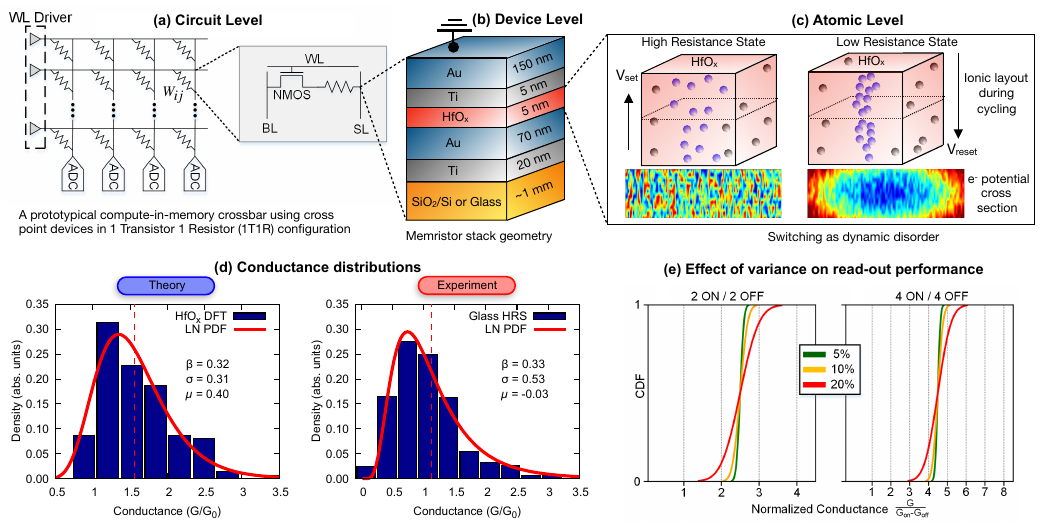}
	\caption{\textbf{Relation between device level variance and circuit-level
		performance. a)} Crossbar architecture with 1-bit word line drivers
		and higher precision ADC's with a NMOS transistor and memristor. In
		this circuit, the word line (WL) and select line (SL) are set to a high
		voltage, and the resulting current along the bit line (BL) is the
		result of the read operation. \textbf{b)} Stack architecture of
		filamentary memristors set length scale at the device level.
		\textbf{c)} Quantum level description of mechanism. The set and reset
		operations generate dynamic potentials for electrons. When the filament
		is fully formed, a high transmission probability path exists.
		\textbf{d)} Measured and simulated conductance fluctuations for HfO$_x$
		showing a log-normal character indicative of phase coherent
		localization. \textbf{e)} Left and Right show total normalized
		conductance of 4 (Left) and 8 (Right) devices with 5, 10, and 20
		percent variance. The black vertical lines represent ADC regions set by
		conductance different of on and on state devices ($G_{\text{on}} -
		G_{\text{off}}$) }\label{fig:one}
\end{figure*}

\section{Overview of compute-in-memory architecture}
Modern computing in general, but deep learning techniques especially, contain a
workload of almost entirely matrix multiplication ($\vec{y} = W
\vec{x}$)~\cite{sze2017efficient}. Neuromorphic hardware is poised to
perform this much more efficiently.  In traditional von Neumann machines, both
the feature data $\vec{x}$ and matrix weights $W_{ij}$ are transported from main
memory to the compute units, where the multiply-and-accumulate (MAC) operations
are performed. After which, the result $\vec{y}$ is transported and written
back into main memory.  In this procedure, the energy cost of reading and
transporting data from memory to logic greatly outweighs the cost of the
MAC~\cite{horowitz20141, chen2017eyeriss}, thus motivating in-memory computing. 

A compute-in-memory paradigm performs the MAC operations in a crossbar
structure using Ohm's law and a non-volatile conductance state. This is often
known as resistive random access memory (RRAM)~\cite{zahoor2020resistive}, but
there are other memories that behave similarly. A circuit-to-atomic level
schematic of the typical (HfO$_x$) RRAM is given in Figure~\ref{fig:one}a-c.
Each element of the matrix $W_{ij}$ is programmed as a conductance in the
crossbar (Figure~\ref{fig:one}a), and each value of the vector $x_i$ is
converted to voltage. The conductance of the active layer
(Figure~\ref{fig:one}b), tuned with external bias, is ultimately set by
controlling the electric potential cross section seen by electrons. These
electrons carry currents along the filaments formed by mobile oxygen vacancies
in disordered HfO$_x$~\cite{gao2019}, as shown in Figure~\ref{fig:one}c.  By
Ohm's law, the current through each RRAM device is proportional to the product
of the programmed conductance $W_{ij}$ and applied voltage $\vec{x}_i$.  By
Kirchhoff’s current law, the resulting currents summed along the columns of the
crossbar are proportional to the product of the matrix and vector, $\vec{y}$.
Using this architecture, the only data transport required is the feature vector
$\vec{x}$ and result $\vec{y}$, whereas a typical von Neumann architecture
would move $\vec{x}$ and $W_{ij}$. Assuming $\vec{x}$ and $W_{ij}$ have the
same dimension $d$, moving only $\vec{x}$ and $\vec{y}$ amounts to a
significant savings of $d(d-1)$ elements.

Although compute-in-memory using the crossbar architecture can greatly reduce
data transport, it faces its own limitations at the device and circuit level.
The two limitations are: (1) the number of states the circuit can read at once
and (2) the number of distinguishable states that can be accurately read from a
column of the crossbar. To read states from the crossbar, an analog-to-digital
converter (ADC) converts the analog current value from the crossbar to a
digital value.  In Figure~\ref{fig:one}a, the end of each column feeds into an
ADC. Modern ADC's are the result of decades of research and can be used to read
thousands of states at relatively high speed~\cite{walden1999analog}. However,
device variance limits the number of distinguishable states that can be
accurately read. Figure~\ref{fig:one}d shows that HfO$_x$ RRAM stacks do have
significant variance. It shows that both the measured and predicted
dimensionless conductance ($g =G/G_o$) of electro-formed HfO$_x$ devices is
distributed log-normally. We will expand in section~\ref{sec:theory} on the
details of the quantum framework that enable this prediction, but at this
point, it is sufficient to say that this is a quantitative model that
also explains the origin of the variability in these devices. 

A log-normally distributed conductance means that problem 2 (clearly
distinguishing states) is exacerbated, as a large proportion of the available
states lie close to each other in value. In Figure~\ref{fig:one}e, we plot the
cumulative distribution function for the total conductance resulting from four
(left) and eight (right) devices.  This figure demonstrates that given enough
variance, the current summed along a column will result in an error with a
probability given by the distribution of the variance of each device. In this
circuit simulation, we use a conductance ratio on par with experiment (see
Figure~\ref{fig:one}d) and sweep five, ten, and twenty percent variance.  In
both cases, ten and twenty percent normalized variance results in erroneous
compute-in-memory results. Although it is a simple analysis, it clearly
demonstrates that increasing the number of devices read will increase the
probability of errors due to accumulated variance, and that any intrinsic
quantum variability is directly tied to circuit level performance. Therefore,
it is desirable to understand the physical origin of the variance in order to
mitigate it. Before laying out the theory however, we will first establish the
well-known conductance properties of filamentary RRAMs.

\section{Conductance variations in filamentary devices}
The conductance of nanoscale, filamentary memristors generally does not follow
a normal distribution. Instead, the log of the conductance is distributed
normally. This is a general feature of these devices, but its origin has been
debated~\cite{lee2015resistive,li2015nano,yi2016nc,degraeve2014hourglass}.
There are phenomenological models that have considered the statistical
properties of the conductance~\cite{karpov2017log}, and heuristic models such
as trap assisted tunneling~\cite{houng1999current} that can recover it, but
none \emph{afford it an origin}. In other words, significant effort has been
devoted to understanding the log-normal conductance in with the goal of
mitigating uncertainty from an engineering perspective, but none have
considered the possibility that a fundamental quantum limit would manifest.
There is, however, this possibility due to the length scales of these devices,
and the stochastic nature of electron transport in disordered systems. 

For example, the many possible configurations of a filament in a thermal
environment are one source of conductance fluctuations~\cite{dattabook}. In
fact, this has been studied quite extensively already, but we do not focus on
that in this work. Another source of variability though is the many possible
paths for electrons to take within any particular filament configuration.
Phase-coherence effects between these different electron paths are known to
give rise to log-normal conductance fluctuations~\cite{abrikosov1981paradox}.
To see this schematically, we revisit figure~\ref{fig:one}c. It depicts a
filamentary device in two arbitrary resistance states. In one resistance state,
the defects form a filament, and in another resistance state, the defects do
not form a continuous filament. In either state, there are many possible defect
configurations, each with a different disorder potential for electrons
(indicated schematically by the heat maps). As a result, these devices can be
considered as dynamically disordered, and transport will be heavily influenced
by phase coherence and localization
effects~\cite{chandrashekar2014quantum,lu2019electronic}. More importantly, if
the transport length $L$ approaches the phase coherence length $L_{\phi}$, the
transmission probability for electrons will approach a universal distribution
given only by the mean conductance--its character being log-normal with a
normalized variance inversely proportional to its
mean~\cite{nieuwenhuizen1995intensity}.

Beyond just a heuristic argument, however, there is additional theoretical and
experimental evidence to suggest quantum effects may play an important role.
The evidence mainly takes the form of thermal and dimensional arguments.  For
example, Basnet et al. have shown that thermally insulating substrates can
improve the performance of HfO$_x$ stacks by decreasing the
variance~\cite{basnet2020substrate}.  Other comprehensive investigations of
HfO$_x$ RRAMs also showed that increasing the temperature of the entire device
from 25$^{\circ}$C to 150$^{\circ}$C reduced the log-normal
tailing~\cite{fantini2013intrinsic}. Beilliard et al. demonstrated that
different stack compositions such as Al$_2$O$_3$/TiO$_{2-x}$ display an
increased variability at cryogenic
temperatures~\cite{beilliard2020investigation}, suggesting these effects are
not even unique to HfO$_x$. All these works are  commensurate with the idea
that increased inelastic  scattering at higher temperature could drive
$L_{\phi}\ll L$, restoring uncorrelated, diffusive transport and normally
distributed conductance.  With regards to dimensionality, Bradley et al. have
shown experimentally that electron injection tends to drive aggregation of
oxygen defects into a quasi-1D filament~\cite{bradley:2015-1,bradley:2015-2}.
The popular trap-assisted tunneling model~\cite{houng1999current} sets the
current proportional to a barrier-tunneling probability, or equivalently, to
the rate of traversing the slowest bridge along a classical percolation path.
Therefore, existing theory and experiment both tend to view the transport in
nanoscale-filamentary memristors as quasi-1D. In spite of this, existing
models are generally semi-classical, even though Funck et al. have shown
that the quantum behavior might be fundamentally different from semi-classical
predictions~\cite{funck2018}.  Thus, both current theoretical models and
experimental evidence suggest phase-coherence effects can play an important
role. Our results indicate that the log-normality of the conductance can in
fact be attributed to quantum disorder effects, and although we use HfO$_x$ as
a model system, these results likely apply to other materials systems. 

\begin{figure}
	\centering
	\includegraphics[width=\columnwidth]{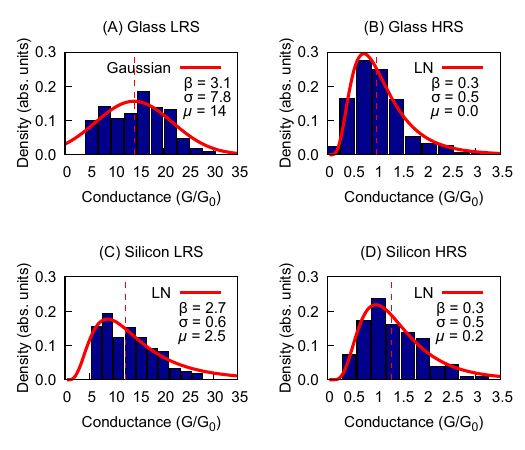}
	\caption{\textbf{Experimentally measured distributions of HfO$_x$ memristors.} 
	The plot shows conductance distributions on different substrates and 
	in both high and low resistance states. The data shows that over 
	many devices, and multiple measurements, a clear statistical 
	distribution emerges that shows log-normal behavior in the conductance, a
	hallmark of phase coherent transport.  
	}\label{fig:two}
\end{figure}

Figure~\ref{fig:two} characterizes the conductance variability in our devices.
The experimental data was obtained by switching the devices between the low
resistance state (LRS) and the high resistance state (HRS) multiple times on
many devices. To switch to the LRS, a positive voltage sweep was applied to the
top electrode from 0 to 1.2 V. After the device was set to the LRS, a voltage
sweep from 0 to 0.1 V was conducted. In post processing, the inverse slope of
this low voltage sweep was calculated to determine the resistance of the
device. A negative voltage sweep from 0 to -1.5 V was used to reset the device
to the HRS. Then another 0 to 0.1 V sweep was used to determine the resistance
of the HRS. Each device was switched between the LRS and HRS ten times to
obtain the cycle-to-cycle variation. This was repeated on 37 devices on a glass
substrate and 36 devices on a thin SiO$_2$/Si substrate.  In each panel, a
histogram of conductance measurements is overlaid with either a log-normal or
Gaussian distribution defined as
\[ 
p(g) = 
\begin{dcases} 
	  \frac{\beta}{g \sqrt{2 \pi \sigma^2}} \text{exp}
	  \left[-\frac{(\text{ln}[g] - \mu)^2}{2\sigma^2}\right] & \text{Log-normal} \\ 
	  \frac{\beta}{\phantom{g} \sqrt{2 \pi \sigma^2}} \text{exp}
	  \left[-\frac{(g - \mu)^2}{2\sigma^2}\right] & \text{Gaussian.} 
\end{dcases}
\] 
We find that in both the HRS and the LRS with a silicon substrate, the
conductance distribution remains log-normal (panels C\&D of
Figure~\ref{fig:two}). In the HRS on glass (panel B of Figure~\ref{fig:two}),
we also find log-normal. In the LRS on glass however (panel A of
Figure~\ref{fig:two}), we find Gaussian or multi-modal, and attribute this to
the difference in thermal conductivity between substrates. These results
clearly show that a log-normal conductance manifests experimentally, and
furthermore, that it can be altered by reducing the heat flow away from the
active layer. 

To summarize the filamentary section, our data, along with the generally
observed log-normal features of filamentary memristors, as well as the
reduction of log-normal tailing with increased temperature, strongly suggests
the origin and behavior of device variance can be described as a quantum wire
with dynamic disorder.  What this means for filamentary RRAM design is that in
addition to variability from stochastic filament dynamics, electron transport
dynamics induce additional variability from interference phenomena. If
dominated by electron phase effects, the conductance variance $\sigma$ is
bounded by the analytical limit of $\sigma \approx 2/(3\langle g
\rangle)$~\cite{nieuwenhuizen1995intensity}. This has several meaningful
consequences.  The first is that to reduce variability and improve RRAM
performance, one must reduce $L_{\phi}$ to be less than the transport length
$L_{\phi} \ll L$.  The second is that the statistical behavior of a single
device with many, bias-induced random filaments is the same as the statistical
behavior of many devices with equally random filaments. This means ergodic
considerations\footnote{The ergodic hypothesis is that after a long enough
time, all the microstates of the system with the same energy will be
accessed equiprobably. It has been widely employed in the study of disorder
effects~\cite{lee1985universal}, and is one of the most basic assumptions
in equilibrium quantum
statistics~\cite{birkhoff1931proof,neumann1932proof,moore2015ergodic}.} can be
taken into account at the circuit level design, i.e., that device-to-device
variance is a manifestation of different disorder distributions, and will be
distributed in the same way as cycle-to-cycle variance.  The third is that the
read-to-read variance would be limited by universal conductance fluctuations
for cryogenic applications if $L_{\phi}$ is larger than the inelastic
scattering length.  Finally, because the variance of synaptic weights is
closely tied to the learning capability of biological and artificial neural
networks~\cite{Buzsaki2014,rajan2006eigenvalue}, the existence of a quantum
source of stochasticity suggests one could leverage it to mimic the limited
precision computation model of biological systems~\cite{Buzsaki2014}. 

\begin{figure*}[ht!]
	\centering
	\includegraphics[width=\textwidth,clip,trim=0 54pt 0 0]{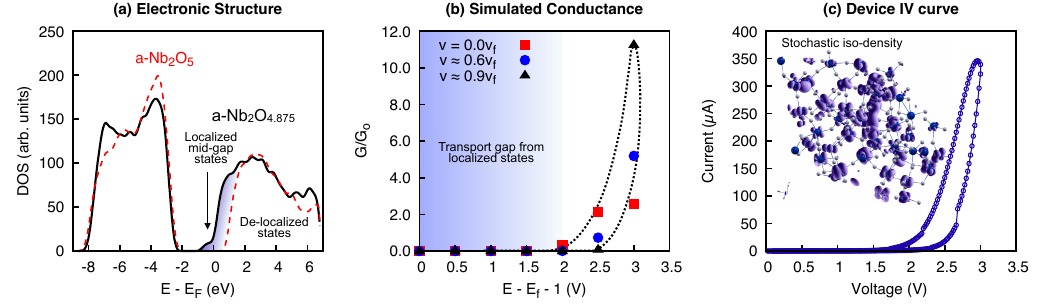}
	\caption{\textbf{Mobility edge switching and hysteresis with dynamic
	disorder. a)} The density of states for amorphous stoichiometric
	(a-Nb$_2$O$_5$) and off-stoichiometric (Nb$_2$O$_{4.875}$) niobium oxide.
	Oxygen deficiency draws out the conduction band and shifts the chemical
	potential by $\sim$ 1 eV.  The shaded region indicates a region of
	strong localization, and the region integrated to produce the isodensity
	surface shown as an inset in part in the rightmost panel. \textbf{b)}
	Simulated conductance with dynamic disorder. By assuming a greater vacancy
	transport (disorder potential redistribution), a single operating bias can
	have multiple conductance states. Each point is averaged over five disorder
	realizations with W = 3 eV. The dotted line is a guide to the eye.
	\textbf{c)} Annular memdiode device IV character. The inset shows the 0.1
	$e^-$/\AA$^3$ isodensity surface  demonstrating the stochastic landscape
	for electrons in establishing a current. }\label{fig:three}
\end{figure*}

\section{Conductance variations in non-filamentary systems}
This section considers the possibility that hysteretic switching in
non-filamentary systems can be equally described within a framework of dynamic
disorder. We are motivated by recent advancements demonstrating that Anderson
localization can be leveraged as a mechanism of purely electronic switching in
silicon~\cite{lu2019electronic}, and by amorphous, electroforming-free, niobium
oxide memdiodes recently developed by Shank et al~\cite{shank2018}. We use
a-Nb$_2$O$_{5-x}$ as a model system to test this hypothesis. The first notable
difference between filamentary and non-filamentary systems is that by
definition, non-filamentary transport cannot be assumed to be quasi-1D.
Therefore, the way quantum conductance variations present themselves could be
totally different. In addition, disorder dynamics are known to be different in
the crystalline phase and the amorphous phase~\cite{Tuller:2017}, so this can
have an effect too. In fact, the introduction of disorder can increase the
ionic conductivity by up to four orders of magnitude in some
materials~\cite{Wohlmuth.Epp.ea:2014, Wohlmuth.Epp.ea:2016}. Considering this, and the
unique propensity of the niobium oxide series to accommodate dynamic
defects~\cite{andersson1966crystallographic,nico2016,kaviani2017}, it is reasonable to
assume that disorder potentials will evolve in three dimensions under high
field. By extension, if the potential felt by electrons in the presence of a
changing ionic environment is dynamic, it is possible that there can be
multiple electron transmission probabilities at a given bias. In this way,
dynamic disorder allows a unified description of filamentary HfO$_x$ and
non-filamentary a-Nb$_2$O$_{5-x}$--the main difference being the dimensionality
of transport. We will show in section~\ref{sec:theory} that the dimensionality,
however, is inherently encoded in the matrix elements of the velocity operator,
so a single framework can treat both cases.

The non-filamentary devices studied here are annular Ni/a-Nb$_2$O$_{5-x}$/Ni
memdiode stacks fabricated on Al$_2$O$_3$ substrates.  They are three-hundred
nanometers thick with a radius of one-hundred micrometers. The normalized
conductance in the high-resistance state is approximately one-hundred times
smaller than HfO$_x$, suggesting stronger localization and 3D transport. By
contrast, the transport direction in the HfO$_x$ RRAMs is approximately five
nanometers, and the transverse direction is sub-nanometer, leading to quasi-1D
transport. Two different memdiode devices were fabricated and tested. A
positive voltage sweep from zero to three volts was applied to the top
electrode of each device, and the subsequent current generated was measured.
This voltage sweep was repeated one-hundred times with a two second delay in
between voltage cycles.  

Figure~\ref{fig:three} summarizes the experimental and theoretical results
pertaining to non-filamentary devices. Panel a gives the first-principles
electronic structure of the a-Nb$_2$O$_{5-x}$ active layer of the memdiode
Computational details are given in the Supplementary Material~\cite{SM}.  Panel b and c
show the theoretical and experimental transport characteristics respectively.
In part a, oxygen deficiency draws out the conduction band and shifts the
chemical potential to a region of finite density of states.  Traditionally,
this would be an indication of metallic transport, but the measurements clearly
show diode-like behavior.  Therefore, the naive, band-theoretical description
has already fallen short.  If, however, we consider these states are localized
due to disorder, then they would give zero contribution to the conductance
until an additional energy is applied to cross the mobility edge. Given the
experimental turn on voltage is near two volts (Figure~\ref{fig:three}c), the
localization picture seems more appropriate than the band picture. Part b of
Figure~\ref{fig:three} shows the simulated conductance as a function of
energy.\footnote{The abscissa is shifted by one electronvolt to connect with
the fact that the electronic structure of oxygen-deficient a-Nb$_2$O$_{5-x}$ is
shifted relative to the stoichiometric counterpart.} The different data points
represent conductance values parameterized by different normalized defect
velocities $v/v_f$ (effectively, how much of the active layer is disordered).
We will provide the complete theoretical details in section~\ref{sec:theory},
but the main result at this stage is that dynamic disorder gives rise to
multiple conductance states at a given energy, even while averaged over several
specific realizations of disorder. More concretely, at $E-E_f-1 = 3$ eV, we
find the conductance is greater with $v=0.9v_f$ than $v=0.6v_f$, but the
opposite is true at $E-E_f-1 = 2.5$ eV. This means an interplay between the
density of states and localization of the wave functions is contributing to the
final conductance. Actually, wave function localization plays an important role
in defining the critical turn-on voltage (details in section~\ref{sec:kubo}).
This variation in the conductance at a given energy can also be interpreted as
a result of the different disorder dynamics in the forward and reverse bias
directions.  Thus, dynamic disorder is a quantum analogue of the dynamic
boundary between regions of high and low dopant concentrations originally
envisioned by Strukov et al~\cite{strukov2008missing}. Part c gives the
measured IV curve.  Comparing the theoretical result in part b and the
experimental result in part c, the similar features indicate dynamic disorder
and localization effects can in fact enable memdiode-like IV behavior. 

To summarize the non-filamentary section, dynamic disorder potentials in 3D
have manifested conductance variations that are quite different than in
quasi-1D structures. Nevertheless, both can be attributed to electron-phase
effects in highly disordered systems. This result suggests that the critical
voltage is dictated by localization, and that the spread in the hysteresis is
related to the dynamic disorder potential.  Because the statistical
distribution and IV characteristics of two very different systems can be
reproduced in a single approach, we resolve dynamic disorder as a viable
quantum framework to treat memristive switching. Given the evidence presented
thus far, we now turn to detailing the theoretical framework, and providing a
complete theoretical justification for its use in this circumstance. 

\section{Theoretical treatment of dynamic disorder} \label{sec:theory}

\subsection{Construction of the Hamiltonian}
A quantum treatment begins with the Hamiltonian. We define it as the sum of
the kinetic and stochastic components as 
\begin{equation} 
	\label{total_hamiltonian} H = H_t + H_s + H_h.
\end{equation} 
Here $H_t$ is the kinetic term, $H_s$ is the substitutional
disorder induced by vacancies or replacements, and $H_h$ is the hopping
disorder arising from deformation of atomic positions, interstitials, etc.
The kinetic term can be expressed as
\begin{equation} \label{quasiparticle_dispersion}
	H_t = \frac{1}{2}\sum_{ij\alpha \beta} (t_{ij}^{\alpha \beta} - 2
		\mu \delta_{ij} \delta_{\alpha \beta}) c^{\dagger}_{i \alpha}
		c_{j \beta}.
\end{equation}
The fermion operators $c^{\dagger}_{i \alpha}$ create (destroy) particles at
site $i$ $(j)$, with orbital and spin character denoted by $\alpha$ $(\beta)$,
and $\mu$ denotes the chemical potential. In the limit of $H_s = H_h = 0$, the
kinetic integrals $t_{ij}^{\alpha \beta}$ are sufficient to determine bulk
transport properties of long-range, ordered solids. They can be calculated
using standard density functional theory techniques and a Fourier
transformation into a real-space, localized basis.\footnote{To obtain the
kinetic integrals of HfO$_x$ and a-Nb$_2$O$_{5-x}$, we have performed
\textit{ab initio} molecular dynamic quenches, and subsequent density
functional theory calculations. Further computational details can be found
in Supplementary Materials~\cite{SM}.} The challenge, however, is that to study the
statistical behavior of many disorder configurations, many, expensive,
molecular-dynamic simulations of large supercells, followed by
density-functional calculations would be required. It is therefore
worthwhile to pursue a cheaper way to model statistical properties, yet
retain (to the extent that it's possible) a completely first-principles
description of the electronic structure. In the spirit of this, the two types
of disorder can easily be parameterized as a perturbation once $H_t$ is
known. Atomic  vacancies or substitutions are diagonal elements in the total
Hamiltonian, and act purely as a local potential as described by
\begin{equation} \label{site_disorder}
	H_{s} = \sum\limits_{ij\alpha \beta} \Theta_{ij}^{\alpha \beta}(W)
	\delta_{ij} \delta_{\alpha \beta} c^{\dagger}_{i \alpha} c_{j \beta}.
\end{equation}
Deformation of atomic positions or interstitials manifest 
as off-diagonal modulation of the kinetic energy density. We make
the approximation that in the amorphous phase, the atomic positions are
randomly perturbed from their equilibrium positions in the ordered state, 
and this effect is written as 
\begin{equation} \label{hopping_disorder}
	H_{h} = \sum\limits_{ij\alpha \beta} (1 - \delta_{ij} \delta_{\alpha \beta})
	\Theta_{ij}^{\alpha \beta}(W) c^{\dagger}_{i \alpha} c_{j \beta}. 
\end{equation}
The matrix elements $\Theta_{ij}^{\alpha \beta}(W)$ appearing in $H_s$ and
$H_h$ are random numbers drawn from a uniform distribution of width $W$
centered at zero. With this distribution, the chemical potential of the
disordered system is the same as the clean system. This is not strictly
required for our argument, but simplifies the analysis.  

With the kinetic term $H_t$ taken from first principles, we need only justify
the disorder potential ($H_s + H_h$) is physical. We achieve this by performing
twenty \emph{ab initio} molecular-dynamic quenches, and calculating the distribution
of the Hartree potential using density functional theory. We then set the
maximum disorder strength $W$ to be less than the full width at half maximum of
this distribution~\cite{SM}.
The advantage of this approach is that we can generate any number of
physically-justified disorder realizations, and gather statistical information
on important metrics such as transport coefficients, the mobility edge, and
wave function localization, for far less computational expense than in
traditional approach. To determine the distribution of the conductance in many,
disordered filaments, this is sufficient. Each filament realizes a random
disorder potential.  The final part however, is to incorporate the dynamic
aspect of a disorder potential. 

To model a dynamic disorder potential in 3D, we make two reasonable assumptions.
First, we assume electrons equilibrate instantaneously compared to ionic
timescales.  Second, we assume defects drift in an applied field. This is an
approximation because diffusion and thermophoresis also contribute to defect
transport~\cite{Basnet.Pahinkar.ea:2020,West.Basnet.ea:2020}. However, this
simple approximation allows us to treat dynamic disorder through a
spatially-dependent probability density. The probability that any given site in
the lattice has a defect, $P_k$, becomes a function of time and applied bias
$P_k \rightarrow P_k(V,t)$. In the linear drift approximation, we can assume
this probability is parameterized at small times by the drift velocity $v$.
Thus, in the absence of an external field, $v=0$, and every site in the lattice
is equally likely to have a defect. But as the external bias is applied,
defects drift, and are less likely reside in a region of $vt$ away from (or
closer to) an electrode (depending on the defect charge and bias polarity). 
Motivated by Strukov et al's boundary idea, we assume
$P_k(V,t) = 0$ if the lattice site is in a region of reduced defects, and
$P_k(V,t) = 1$ in the other region. In principle, the nature of the probability
function can be adapted to include any number of effects, and if it were exact,
then the description would be exact.  This approach effectively redefines
Equation~\ref{total_hamiltonian} to become bias and time dependent.  The
coupled, system-environment Hamiltonian $H(V,t)$ can be written (ignoring
$H_h$) as 
\begin{equation} \label{time_dependent_disorder}
	H(V,t) = H_t + \sum\limits_{ij\alpha \beta} \Theta_{ij}^{\alpha \beta}(W) 
	P_{ij}^{\alpha \beta}(V,t)
	\delta_{ij} \delta_{\alpha \beta} c^{\dagger}_{i \alpha} c_{j \beta}.
\end{equation}

\begin{figure}
	\centering
	\includegraphics[width=0.9\columnwidth]{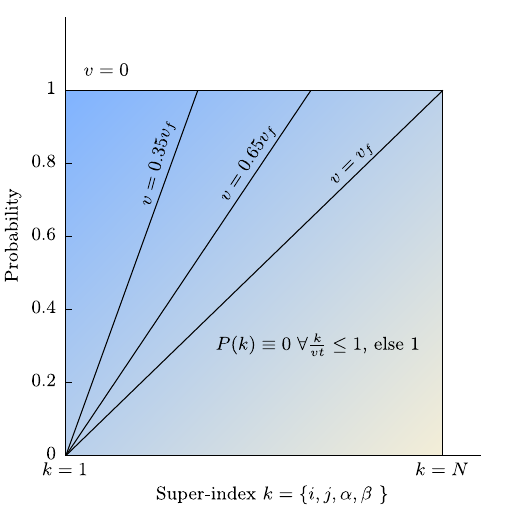}
	\caption{\textbf{Characterizing a dynamic distribution of disorder.} The 
	plot shows the probability distribution of having a defect is dependent 
	on the ionic velocity $v$.} 
	\label{fig:dyna_prob}
\end{figure}

Figure~\ref{fig:dyna_prob} shows the linear probability distribution
implemented as a function of Hamiltonian super-index $k$ (covers site and
orbital indices). This is used in simulating the conductance of
a-Nb$_2$O$_{5-x}$ (Figure~\ref{fig:three}b).  The central quantity is the
defect velocity $v$.  If $v=0$, there is no motion of defects, and every site
is equally likely to have a defect. If $v \neq 0$, then the probability for a
site to have a defect changes proportional to $v$.  We calibrate a limiting
velocity $v_f$ by the equilibration times for IV curves in our
a-Nb$_2$O$_{5-x}$ devices. For example, if it takes one-hundred seconds for the
current to stabilize in a three-hundred nanometer thick sample, we may assume
the slowest moving ions are traveling three nanometers per second. This is
reasonable for good ionic conductors~\cite{Steinruck.Takacs.ea:2020}.  This
defines the shallowest slope of our probability curve ($v = v_f$), but if ions
have a smaller propensity to move, we can set the slope to a fraction of $v_f$.

To implement this approach, one chooses an arbitrary measurement time and a
velocity. This defines the disorder potential. Then, the Hamiltonian is
constructed, exactly diagonalized, and the wave functions are used to
determine the transport coefficients (linear response theory details in
section~\ref{sec:kubo}). It is worth mentioning this approach is \emph{not} a
simulation of the ionic motion itself, rather, it enables a
quantum-statistical description of the memristive transport. The time dynamics
of the potential we have employed here is only an approximation.
Nevertheless, by taking a linear one, we have found the spread in the
conductance at a given bias is in line with the experimental result.

\subsection{Transport in the presence of disorder: Validity of Kubo expression}
\label{sec:kubo}
\begin{figure*}[ht!]
	\centering
	\includegraphics[width=0.8\textwidth]{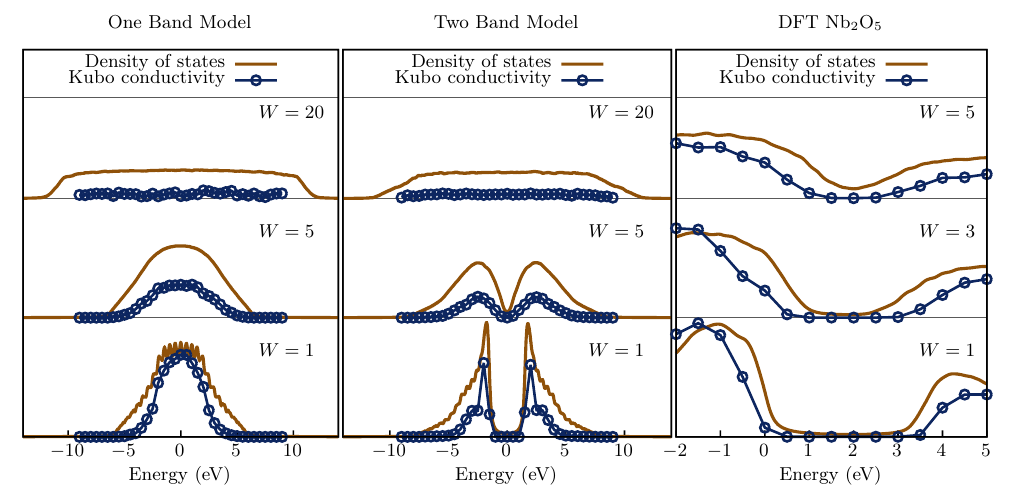}
	\caption{\textbf{Quantifying in-gap contribution to the conductivity.} The
		dotted lines plot the finite size Kubo conductivity as a function of
		energy at 300K in a single band model (left), the two-band model
		(center) and  the first principles Nb$_2$O$_5$ (right). The solid lines
		show the density of states on the same dependent axis with globally arbitrary, 
		but internally relative vertical scale.}\label{fig:conductivity_vs_energy}
\end{figure*}
Thus far, we have not established a theoretical approach to determine the
transport coefficients in dynamic disorder environments. We use the finite-size
implementation of the Kubo conductivity tensor defined as~\cite{allankubo2007}
\begin{equation} \label{finite_size_kubo_formula}
	\sigma_{\alpha \beta}(\mu, T)
	\equiv 
	\frac
		{-i \hbar e^2}
		{N} 
	\sum_{n n'}
	\frac 
		{\Delta f_{nn'}}
		{\varepsilon_{n} - \varepsilon_{n'}} 
	\frac
		{\langle n | v_{\alpha} | n' \rangle
		 \langle n' | v_{\beta} | n  \rangle}
		{\varepsilon_{n} - \varepsilon_{n'} + i \eta}.
\end{equation}
In Equation~\ref{finite_size_kubo_formula}, $N$ is the number of unit cells,
$|n\rangle$ is an eigenstate of the total Hamiltonian $H$ with eigenenergy
$\varepsilon_n$, the velocity operator is $v$, and $\Delta f_{nn'}$ represents
the occupation difference between the state $n$ and $n'$. $\Delta f_{nn'}$ 
carries implicit dependence on temperature and chemical potential through the 
Fermi function. The matrix elements of the velocity operator are defined by
Equation~\ref{velocity_operator} where $r_i$ is the real space position of
state $i$ and $\hat{e}_{\alpha}$ is the unit vector along direction $\alpha$.
\begin{equation} \label{velocity_operator}
	i \hbar v_{ij}^{\alpha} = 
	(r_i - r_j) 
	\cdot
	\hat{e}_{\alpha} H_{ij}
\end{equation}

To establish the validity of the finite-size Kubo formula for applications to
real material systems near an Anderson transition, we benchmark against model
systems with a known localization transition. We choose a one- and two-band,
nearest-neighbor, tight-binding model with cubic symmetry and 1331 unit cells.
The single-band Hamiltonian has a known localization transition at
$W=16.5t$~\cite{Markos:2006}, but because we are primarily interested in gapped
systems, we also benchmark the two-band model. The dispersion in the one band
case is simply defined as
\begin{equation}
	E(k) = -2t(\cos(kx) + \cos(ky) + \cos(kz)).
\end{equation}
It is common to set the hopping parameter $t$ in this one-band model to unity.
The bandwidth is then 12$t$.  In the two-band model, there are two on-site
degrees of freedom with energy $\pm \epsilon$. In the Hamiltonian the only
non-zero hopping elements $t_{ij}^{\alpha \beta}$ are defined as
\begin{equation}
	\begin{split}
		t^{i = j}_{\alpha, \alpha} &= - \epsilon \\
		t^{i = j}_{\alpha, \beta}  &= e^{-2 | \mathbf{r_i} - \mathbf{r_j} | } \\
		t^{i = j}_{\beta, \alpha}  &= e^{-2 | \mathbf{r_i} - \mathbf{r_j} | } \\
		t^{i = j}_{\beta, \beta}   &= + \epsilon.
	\end{split}
\end{equation}
Using these expressions, we can investigate the electronic structure and
wavefunction localization as a function of disorder $W$ across the transition
in known systems, and compare that to the density functional theory (DFT)
result for Nb$_2$O$_{5-x}$. The details for obtaining the DFT result are given
in the Supplementary Materials~\cite{SM}.

Figure~\ref{fig:conductivity_vs_energy} shows the Kubo response as a function
of energy for various strengths of disorder in the various systems overlaid
with the density of states. The solid lines show the density of states, and the
dotted lines show the Kubo conductivity. As expected, we see the one-band and
two-band models undergo a transition to a completely transportless phase with
increasing disorder strength. This demonstrates the finite size Kubo expression
is in fact capable of capturing Anderson localization. The evidence is that we
see a finite density of states, yet zero conductivity inside the Anderson
localized phase ($W=20$). The implications for oxygen-poor a-Nb$_2$O$_{5-x}$
are shown in the right panel of Figure~\ref{fig:conductivity_vs_energy}, where
at $W=5$ eV, the highly localized in gap states give zero contribution to the
linear response. The importance of the matrix element effects are readily
apparent across all systems because the finite density of states across the
chemical potential means the occupation factor $\Delta f_{nn'} /
(\varepsilon_{n} - \varepsilon_{n'})$ in the Kubo expression is positive, but
the phase coherence forms a transport gap. This is only possible if the
velocity matrix elements go to zero.  This analysis shows that for pairs of
wave functions with similar thermal occupation, or small eigenstate coupling,
the contribution to the finite size Kubo conductivity will be small, and that
interference effects are captured by the matrix elements of the velocity
operator. Because the time scale for the ionic drift is usually much slower
than that for the electronic response, the electronic structure will be
established immediately after the oxygen vacancies reach new locations.  Given
these pieces of evidence, the Kubo formula is a valid approach to compute the
conductivity at each snapshot in time.

%and this zero-conductivity
%region coincides with increasing Gini coefficient from
%Figure\ref{fig:gini_demonstration}.  This is evidence the Gini coefficient and
%Kubo expression are viable tools to understand the transport properties
%directly from the character of the wavefunctions. Therefore, predicting a
%disorder-induced transport gap in Nb$_2$O$_{5-x}$ is important for memdiode
%engineering because the energy necessary to drive electrons across the active
%layer will be defined by the mobility edge instead of the traditional
%conduction band edge. Additionally, it suggests the disorder strength could be
%a tuning parameter for the threshold voltage and diode variation. 

\begin{figure*}[ht!]
	\centering
	\includegraphics[width=0.8\textwidth]{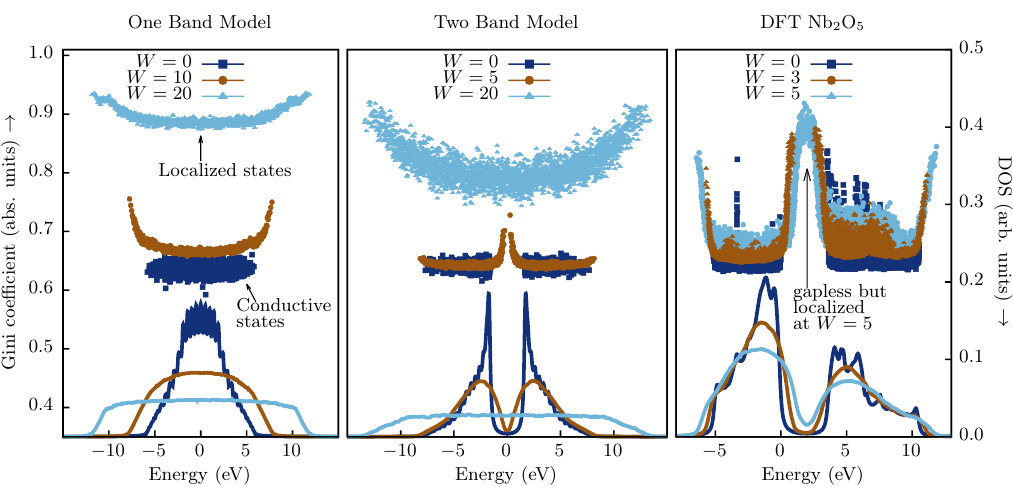}
	\caption{\textbf{Modeling the population of in-gap states and their localization.}
		The lines plot the density of states as a function of disorder strength
		$W$, and the dots represent the Gini coefficient. The left panel is a
		single-band model. The center panel is an arbitrarily gapped, two-band 
		model. The right panel is the density functional result for a 50-band 197-unit
		super-cell of Nb$_2$O$_5$. $W$ specifies the width of the box distribution in
		electron volts}\label{fig:gini_demonstration}
\end{figure*}

\subsection{Characterization of localized wave functions}
To determine if the mid-gap states contribute to transport in the presence of
localization effects, we handcraft a metric of localization focused on
transport.  In some sense it is analogous to the well known inverse
participation ratio~\cite{thouless1974}, but considers not only wave function
extent, but also virtual coupling of all states in the system. To do so, we
first define the object
$\phi^{mn}$ as
\begin{equation} \label{overlap_vector}
	\phi^{mn} = \sum_{r} \big| \langle m | r \rangle \langle r | n \rangle \big|^2.
\end{equation}

If $\{|n\rangle\}$ and $\{|m\rangle\}$ represent eigenvectors of the total
Hamiltonian $H$ in Equation~\ref{total_hamiltonian}, then Eqn~\ref{overlap_vector}
can interpreted as a collection of the probabilities for states to couple
through the eigenspace. We would like to characterize the dispersion of
$\phi^{mn}$, so we can define the Gini coefficient~\cite{ceriani2012} of the
$n^{th}$ eigenstate with energy
$\varepsilon_n$ as
\begin{equation} \label{gini_dispersion}
	g(\varepsilon) = \sum_n \chi^n \delta(\varepsilon - \varepsilon_n),
\end{equation}
where
\begin{equation} \label{gini_coefficient}
	\chi^n \equiv \big[ 2 N \sum \limits_{m} \phi^{mn} \big]^{-1}
	\sum \limits_{m m^{\prime}} \big|\phi^{mn} - \phi^{m^{\prime}n} \big|.
\end{equation}

In this formulation, $g \in [0,1]$, and will describe the distribution of an
electron's propensity to change state.  If this state is characterized by $g =
0$, then it has equal propensity to transition to any other state in the
system. This is not the case in real systems, so in practice the Gini
coefficient of active states lies in the middle of the measure. On the other
hand, if $g = 1$, the state cannot transition to any other state, and this is a
physically realizable situation by many mechanisms.  Therefore, states with a
Gini coefficient of one, will effectively be silent, as they do not couple
through the eigenspace to any other state. This approach may prove to
generalize beyond the inverse participation ratio in characterizing transport
properties because there are many transport mechanisms facilitated by spatially
local wave functions~\cite{baranovski2006}.  While we have only considered
electronically coupled states here, in principle this could be extended to
include states coupled by other mediators, for example phonons in variable
range hopping situations. When connected to \textit{ab initio} simulations of
disordered systems, this is a simple yet powerful method to gauge the mobility
edge in real materials.  

%However, we still would like to know how the interference
%localized states contribute to the conductivity, and for this we employ the
%finite size Kubo formula. 

%This calculation also serves to validate the Gini vector as a metric of
%localization. Because the single-band case has a known localization transition
%at $W=16.5t$~\cite{Markos:2006}, the behavior of the finite-size Kubo formula
%and the Gini vector across this transition can be tested simultaneously. 

%In the three-dimensional, single-band
%case, there is a well known transition to a fully transportless phase when the
%disorder strength $W$ reaches $16.5/t$~\cite{Brandes.Kettemann:2003}, with $t$
%setting the kinetic energy scale.  This is reflected in the left panel by the
%Gini coefficient tending towards unity inside the localized phase $W=20$.
%Similar trends are seen in the two band model shown in the center panel of
%Figure~\ref{fig:gini_demonstration}.

The density of states and Gini coefficient as a function of disorder are shown
in the panels of Figure~\ref{fig:gini_demonstration}. The density of states is
shown by the solid lines and the Gini coefficient by the points. In all three
panels, the abscissa is absolute and physical. The ordinate for the Gini
coefficient is absolute, but the ordinate for the density of states is only
internally relative and arbitrary. The leftmost panel shows the single band
model; the center panel the two-band model; and the right panel shows the DFT
result for Nb$_2$O$_5$.  The single and two band models shown in
Figure~\ref{fig:gini_demonstration} are again used as a benchmark.  For zero
disorder, the Gini coefficient is uniform, and the insulating gap is clearly
visible. As the disorder is increased to $W=3$, the Gini coefficient spikes for
states near the chemical potential, then quickly decays at higher energy. This
demonstrates that in principle, we can resolve a mobility edge. The right panel
shows the first principles result for Nb$_2$O$_5$.  At $W=0$ eV, there is an
insulating gap in the density of states and the Gini coefficient. As the
disorder is increased to $W=3$ eV, the gap begins to close but remains finite,
and the Gini coefficient of the in-gap states spikes. At $W=5$ eV, states
completely span the gap. In clean systems, this would be an indication of
metallic behavior. However, the extent to which states inside the gap couple to
all other states is greatly reduced, indicated by a tendency of the Gini
coefficient towards unity. This situation greatly reduces the number of
conducting channels available, and indicates phase-coherent localization of these 
mid gap states. 

To summarize the theoretical section, development of the dynamic disorder
potential, a viable implementation of the linear response, and a metric of
localization represents a framework to understand quantum-statistical effects
in memristive computing architectures. Predictions in filamentary and
non-filamentary systems have been presented in Figures~\ref{fig:one},
\ref{fig:two}, and \ref{fig:three}--they compare well with experiment. Figures
\ref{fig:gini_demonstration} and \ref{fig:conductivity_vs_energy} demonstrate
that we can resolve localized wavefunctions in our approach, and that the Kubo
formula is a viable tool to model the linear response. One framework gives a
quantum-statistical origin to the log-normal conductance fluctuations in
filamentary RRAMs, and enables quantitative predictions. It also exposes the
role of localization in the critical voltage and hysteresis in non-filamentary
systems. With this in mind, we now discuss the general implications of this
framework for in-memory computing. 

\section{General impact for compute-in-memory architectures}
With the evidence presented thus far, some rather general and striking aspects
of the agency of neuromorphics to transcend Moore's law emerge.  These general
aspects originate in the behavior of nanoscale memristors. They are:
\begin{enumerate}[itemsep=0pt,topsep=4pt]
		\item Basic quantum properties of disordered systems at the nanoscale
			are an \emph{immutable} source of variability. 
		\item The ergodic hypothesis
			ensures that device-to-device variability is the same as
			cycle-to-cycle variability, as each are just different
			instantiations of disorder.
		\item The logarithm of the conductance, not the resistance,
			obeys the central limit theorem.
\end{enumerate}
So while the classical conception of neuromorphics is to accelerate matrix
operations using a \emph{well-defined} state, there is an additional, and
possibly more important ability of a truly neuromorphic system; that is the
ability to fall into non-deterministic states.  Any machine, no matter how
complex or multi-variate, can hardly be considered intelligent if its products,
possibly ad infinitum, are yet deterministic. 

Therefore, the promise of neuromorphics lies not only in the ability to
circumvent the von Neumann bottleneck, but also in the ability to leverage
quantum non-determinism for truly intelligence hardware designs. Future efforts
can now be directed towards learning algorithms incorporating the
statistical nature of logical nodes. For example, one may envision a simple
neural network where each node's state is not known precisely, but rather, is
drawn from a log-normal distribution with variance inversely proportional to
the running average of its activations. This is how the brain itself
works~\cite{Buzsaki2014}, and as such, a machine of this type may have greater
generalizability due to the finite probability of each of its nodes to take on
values far from their means. In this sense, what has thus far been considered a
hindrance, is realized to actually be a utility, and is rooted in the quantum
transport properties of nanoscale disordered systems. 
 
\section{Conclusion} 
We have presented a unified description of the switching characteristics and
intrinsic variability of two very different classes of memristive devices,
suggesting design paradigms and connections to the fundamental processes of
bio-mimetic learning. Our framework's ability to reproduce salient experimental
signatures shows that quantum interference phenomena are directly linked to
circuit level performance with implications for endurance, reliability, and
scaling of neuromorphic hardware. Although we have demonstrated the close
relation between oxygen deficiency, disorder, and memristive response in
electro-forming free a-Nb$_2$O$_{5-x}$ and filamentary a-HfO$_x$, the minimal
ingredients for this physics can be found in many systems, requiring only
dynamic disorder.  In fact, we suspect that memristive hysteresis should be
found in practically all disordered systems as long as there are nearby
meta-stable configurations that can be accessed with an impulse. The
understanding  gained from circuit realizations of intelligence may even
provide insight into the functioning of human brain.

\begin{acknowledgments}
This work was supported by the Air Force Office of Scientific Research
Multi-Disciplinary Research Initiative (MURI) entitled, “Cross-disciplinary
Electronic-ionic Research Enabling Biologically Realistic Autonomous Learning
(CEREBRAL)” under Award No. FA9550-18-1-0024 administered by Dr. Ali Sayir. \\
\end{acknowledgments}

%\section{Data Availability}
%The data for this study can be made available upon reasonable request.

%\section{Code Availability}
%The code for this study can be made available upon reasonable request.

\section{Author Contributions}
C.N. Singh, L.F.J. Piper, and W.-C. Lee conceived the study of conductance distributions. 
C.N. Singh designed and performed the theoretical analysis of the conductance fluctuations.
K. T. Butler performed the molecular dynamic simulations. 
B. A. Crafton and A. Raychowdury conducted the circuit analysis. 
A. S. Weidenbach and W. Alan Doolittle fabricated and performed experiments for 
Nb$_2$O$_{5-x}$ devices, M. P. West and E. M. Vogel for HfO$_x$. 
C.N. Singh, L.F.J. Piper, A. H. MacDonald, and W.-C. Lee analyzed the
experimental data and theoretical simulations. 
C.N. Singh, L.F.J. Piper, and W.-C. Lee wrote the manuscript with inputs from all authors. 
C.N. Singh, L.F.J. Piper, and W.-C. Lee ensured the clarity of the manuscript.

\section{Competing Interests}
The authors declare no competing interests.

~\nocite{Momma.Izumi:2011,Schwarz.Blaha.ea:2002,Mostofi.Yates.ea:2014,Nico.Monteiro.ea:2016,Wahila.Paez.ea:2019,Blochl:1994,Kresse.Hafner:1993,Perdew.Ruzsinszky.ea:2008,Krukau.Vydrov.ea:2006}


\begin{thebibliography}{75}
\bibitem [{SM()}]{SM}%
  \BibitemOpen
  \href@noop {} {\bibinfo {title} {See {S}upplemental {M}aterial at [{URL} will
  be inserted by publisher] for details of the first-principles modeling, the
  wannierization, the variation of the hartree potential, and the scaling
  analysis. {T}he {S}upplemental {M}aterial includes references
  [67-75].}}\BibitemShut {Stop}%
\end{thebibliography}
\end{document}